# The NRGTEN Python package: an extensible toolkit for coarse-grained normal mode analysis of proteins, nucleic acids, small molecules and their complexes

Olivier Mailhot[1,2] & Rafael Josef Najmanovich[1,*]

[1] Department of Pharmacology and Physiology, Faculty of Medicine, Université de Montréal

[2] Institute for Research in Immunology and Cancer (IRIC), Faculty of Medicine, Université de Montréal

[*] To whom correspondence should be addressed: rafael.najmanovich@umontreal.ca

**Summary:** Coarse-grained normal mode analysis (NMA) is a fast computational technique to study the dynamics of biomolecules. Here we present the Najmanovich Research Group Toolkit for Elastic Networks (NRGTEN). NRGTEN is a Python toolkit that implements four different NMA models in addition to popular and novel metrics to benchmark and measure properties from these models. Furthermore, the toolkit is available as a public Python package and is easily extensible for the development or implementation of additional NMA models. The inclusion of the ENCoM model (Elastic Network Contact Model) developed in our group within NRGTEN is noteworthy, owing to its account for the specific chemical nature of atomic interactions. This makes possible some unique predictions of the effect of mutations, such as on stability (via changes in vibrational entropy differences), on the transition probability between different conformational states or on the flexibility profile of the whole macromolecule/complex (to study allostery and signalling). In addition, all NMA models can be used to generate conformational ensembles from a starting structure to aid in protein-protein, protein-ligand or other docking studies among applications. NRGTEN is freely available via a public Python package which can be easily installed on any modern machine and includes a detailed user guide hosted online.
**Availability and implementation:** https://github.com/gregorpatof/nrgten_package/
**Contact:** rafael.najmanovich@umontreal.ca

## Introduction

Coarse-grained normal mode analysis (NMA) models, also known as elastic network models (ENMs), have been used in the past to study protein and nucleic acid dynamics. The advantages of ENMs are their low computational cost, the easy access to long timescales and their robustness to small variations in the starting structure. Moreover, contrary to time consuming all-atom NMA performed using classical forcefields, the starting structure does not need to be minimized in order to compute its normal modes using an ENM (Dykeman & Sankey, 2010). Despite their usefulness, most ENMs use the backbone geometry alone to determine the normal modes and are thus insensitive to the chemical nature of the amino acids/nucleotides/ligands considered. This was the motivation behind the development of ENCoM (Elastic Network Contact Model), an ENM developed by the Najmanovich group and sensitive to atomic interactions of the studied macromolecules without much added computational cost (Frappier & Najmanovich, 2014). This all-atom sensitivity allows ENCoM to predict the effect of mutations on protein stability (Frappier

& Najmanovich, 2015) and to date no other ENMs, to the best of our knowledge, can perform such predictions. ENCoM is freely available as open-source software. Whereas a web server was previously accessible (Frappier, Chartier, & Najmanovich, 2015), due to circumstances outside of our control, the server is no longer available. Moreover, the original ENCoM implementation was difficult to install, its use was complicated by the calling of multiple executables and the generation of intermediate files, the source code was not extendable and did not permit the calculation of transition probabilities. We thus present the Najmanovich Research Group Toolkit for Elastic Networks (NRGTEN), an open-source Python package that implements ENCoM and other popular ENMs along with commonly used metrics and new measures for ENMs, most notably transition probabilities. NRGTEN is thoroughly documented in an online user guide covering basic as well as more advanced uses of the software. It can be installed using a single terminal command on macOS and Linux systems and requires only a few additional steps on Windows.

**Implementation**
NRGTEN currently implements the following ENMs: the Anisotropic Network Model (ANM) (Atilgan et al., 2001), a modified version of ANM called "parameter-free" (Yang, Song, & Jernigan, 2009), the Generalized Spring Tensor Model (STeM) (Lin & Song, 2010) and ENCoM (Frappier & Najmanovich, 2014). NRGTEN also allows the quick implementation of other ENMs by simply extending the ENM Python class, freeing the user from the burden of parsing PDB files, building the internal representation and hand-coding the solver of the Hessian matrix. NRGTEN also implements various existing and new metrics (see below and the online user guide for more details). The following uses all mention ENCoM as the default model, however any ENM could be used in its place. Keep in mind that applications involving the effect of mutations would have very poor if not random results if computed using sequence agnostic ENMs like ANM or STeM. All the examples below are documented with step-by-step instructions in the online user guide (https://nrgten.readthedocs.io).

**Typical uses**
**Dynamic Signatures:** ENCoM can be used to generate what we call dynamic signatures, which are vectors of predicted fluctuations in local flexibility at each position in the macromolecule (each amino acid in the case of proteins). These dynamic signatures can be used to cluster sequence variants of a protein according to their flexibility patterns or to screen for a given effect (e.g., rigidification of a domain) among *in silico* mutants. We recently applied such dynamic signatures to SARS-CoV-2 Spike protein variants to predict their infectivity (Teruel, Mailhot, & Najmanovich, 2020).

**Conformational ensembles:** A conformational ensemble generated using a combination of the first few low-frequency normal modes can be easily created as a multi-model PDB file. Such conformational ensembles generated from ENMs can be used to simulate protein flexibility with applications in small-molecule (Kurkcuoglu & Doruker, 2016) and protein-protein docking as well as to fit high-resolution structures into low-resolution cryo-EM maps (Tama, Miyashita, & Brooks, 2004).

**Vibrational entropy differences:** These can be evaluated between conformational states of a protein or between sequence variants. Past studies have shown that changes in rigidity as a result of mutations affect thermal stability (Frappier & Najmanovich, 2015).

**Transition probabilities between states:** We introduce here a new measure that calculates the transition probabilities between conformational states. Interestingly, transition probabilities calculated with ENCoM can be used to predict the equilibrium occupancies between conformational states and predict the effect of mutations on infectivity for variants of the SARS-CoV-2 Spike protein (Teruel et al., 2020).

**Advanced uses**
**Adding residues or ligands to ENCoM:** The sensitivity of ENCoM to the chemical nature of residues comes from the evaluation of surface areas in contact between all heavy atoms. The eight atom types from (Sobolev, Wade, Vriend, & Edelman, 1996) need to be assigned to all heavy atoms which are to be considered by the model. This has already been done for amino acids, standard nucleic acids (dA, dC, dG, dT, A, U, C, G) and common modified nucleotides. However, a user wanting to use ENCoM on a protein-ligand complex, or on a macromolecule with chemical modifications, can easily define the atom types of the atoms within the non-standard chemical groups by supplying simple additional input files (see user guide for detailed instructions). Therefore, NRGTEN can be used to compare apo and holo states of a macromolecule as well as post-translational modifications and nucleic acids with non-standard nucleotides.

**Metrics for benchmarking ENMs:** Various important metrics have emerged over the years to test the reliability of ENMs. The following metrics are included as part of NRGTEN: the overlap between the normal modes and a given conformational change (Marques & Sanejouand, 1995), the cumulative overlap for a given number of normal modes (Zimmermann & Jernigan, 2014), principal components analysis (PCA) from an ensemble of structures (Hayward & Groot, 2008), root-mean-square inner product (RMSIP) between a given number of normal modes and principal components (Leo-Macias, Lopez-Romero, Lupyan, Zerbino, & Ortiz, 2005).

**Implementing custom ENMs:** The object-oriented architecture of NRGTEN enables its use as a platform to implement and test new ENMs with a minimal amount of additional code required. In particular, users can define the level of coarse graining with control over the definition of nodes in the elastic network (e.g., using more nodes for larger amino acids in proteins).

**Conclusions**
We have developed the Najmanovich Research Group Toolkit for Elastic Networks (NRGTEN), an expanded, extensible and user-friendly toolkit for coarse-grained elastic networks as a Python package available at https://github.com/gregorpatof/nrgten_package/. NRGTEN decreases the barriers to employing NMA methods in structural bioinformatics studies by enabling the easy installation and use of various elastic networks, most prominently ENCoM, a sequence-sensitive elastic network model. NRGTEN is documented in great detail in an online user guide (https://nrgten.readthedocs.io) with both non-technical and expert users in mind. NRGTEN is available on all major platforms and can be installed in just a few steps on Windows and a single step on macOS and Linux. The package allows the easy computation of normal modes for macromolecules and their complexes and provides functions to compute dynamic signatures, conformational ensembles, vibrational entropy differences and transition probabilities between states. NRGTEN also allows for the easy addition of new chemical entities such as modified residues/nucleotides or ligands, includes benchmarks commonly used in the ENM field and enables the fast and easy implementation of novel ENMs.

**Funding**
OM is the recipient of an FRQ-NT PhD fellowship. RN is FRQ-S Senior Research Fellow and a member of Proteo and RQRM.